\documentclass[aps,showpacs,twocolumn,showkeys,amsmath,amssymb,floatfix,nofootinbib,preprintnumbers]
{revtex4}
\usepackage{bm}
\usepackage{array}

\def\de{\delta^{\vphantom{1}}}
\def\bde{{\bar\delta}}
\def\qq{{q\bar q}}
\def\QQ{{Q\bar Q}}
\def\dep{{\delta^{\prime\vphantom{1}}}}
\def\B{B^{\vphantom{1}}}
\def\bt{{\bar\theta}}
\def\hf{{\textstyle{\frac 1 2}}}
\def\h3{{\textstyle{\frac 3 2}}}

\begin{document}
 
\title{Spectroscopy of Exotic Hadrons Formed from Dynamical Diquarks}

\author{Richard F. Lebed}
\email{richard.lebed@asu.edu}
\affiliation{Department of Physics, Arizona State University, Tempe,
Arizona 85287-1504, USA}

\date{September, 2017}

\begin{abstract}
The dynamical diquark picture asserts that exotic hadrons can be
formed from widely separated colored diquark or triquark components.
We use the Born-Oppenheimer (BO) approximation to study the spectrum
of states thus constructed, both in the basis of diquark spins and in
the basis of heavy quark-antiquark spins.  We develop a compact
notation for naming these states, and use the results of lattice
simulations for hybrid mesons to predict the lowest expected BO
potentials for both tetraquarks and pentaquarks.  We then compare to
the set of exotic candidates with experimentally determined quantum
numbers, and find that all of them can be accommodated.  Once decay
modes are also considered, one can develop selection rules of both
exact ($J^{PC}$ conservation) and approximate (within the context of
the BO approximation) types and test their effectiveness.  We find
that the most appealing way to satisfy both sets of selection rules
requires including additional low-lying BO potentials, a hypothesis
that can be checked on the lattice.
\end{abstract}

\pacs{14.40.Rt, 14.40.Pq}

\keywords{Exotic hadrons, quarkonium, diquarks, Born-Oppenheimer
approximation}
\maketitle


\section{Introduction}
\label{sec:Intro}

In a period of less than 15 years, the number of observed heavy-quark
exotic hadron candidates has grown from none to over
30~\cite{Lebed:2016hpi,Esposito:2016noz,Ali:2017jda}.  Even so, the
nature of the substructure of these novel states remains hotly
disputed.  In addition to the possibility that some of the neutral
exotics are heavy quark-antiquark ($\QQ$) hybrid
states~\cite{Meyer:2015eta}, a variety of multiquark options have been
advocated: hadronic molecules $(Q {\bar q}^\prime)(\bar Q q)$ of
color-singlet hadrons, including kinematic enhancements due to the
proximity of hadronic thresholds~(reviewed in~\cite{Guo:2017jvc});
hadroquarkonium~\cite{Voloshin:2007dx,Dubynskiy:2008mq}, in which the
$\QQ$ pair forms a compact core surrounded by a larger light-quark
${\bar q}^\prime q$ wave function; and diquark models~(most notably,
in Ref.~\cite{Maiani:2004vq}), in which the quark (and antiquark)
pairs form close associations through the attractive color ${\bf 3}
\otimes {\bf 3} \to \bar {\bf 3}$ and $\bar {\bf 3} \otimes \bar {\bf
  3} \to {\bf 3}$ channels to form quasi-bound diquarks, $\delta
\equiv (Qq)_{\bar{\bf 3}}$ and $\bar \delta \equiv (\bar Q \bar
q^\prime)_{\bf 3}$, respectively.

The dynamical diquark {\em picture}~\cite{Brodsky:2014xia} is a
physical paradigm in which some of the light quarks $q, \bar q^\prime,
\ldots$ created in the production process of a heavy quark-antiquark
pair $\QQ$ not only coalesce into diquarks through the color mechanism
just described, but achieve a substantial spatial separation by virtue
of recoils achieved through the large energies available in processes
such as $b \to c$ decays or collider events.  Originally posited as a
natural mechanism for creating tetraquark states that remain strongly
bound despite their large spatial extent ($> 1$~fm) due to color
confinement of the diquark-antidiquark pair, the picture can easily be
extended to pentaquarks and beyond~\cite{Lebed:2015tna}, by using the
successive accretion of additional quarks through the color-triplet
channel attraction.  For example, pentaquarks can be interpreted as
triquark-diquark $\bar \theta \delta \! \equiv \! [\bar Q(q_1
q_2)_{\bar{\bf 3}}]^{\vphantom{(}}_{\bf 3} (Q q_3)_{\bar{\bf 3}}$
states.  The substantial relative strength of the diquark
color-triplet attraction compared to the quark-antiquark color-singlet
attraction (which is a factor of $\frac 1 2$ at short distances)
suggests that diquark formation should be a common feature of hadronic
processes; for example, a simple treatment of a collection of quarks
and antiquarks as a static ideal gas predicts diquark attraction to be
the dominant interaction $O(10\%)$ of the time~\cite{Lebed:2016epe}.

In order for this picture to be physically meaningful, the diquarks
must be somewhat spatially compact and achieve some reasonable spatial
separation, so that the state may exhibit some distinctive physical
signature different from that of other structures, such as compact
tetraquarks or hadronic molecules.  ``Reasonable'' in this sense means
that the bulk of the wave functions of the distinct diquarks do not
significantly overlap.  Alternatively, it is worth noting that
diquarks have instead been considered as long-distance correlated
quark pairs~\cite{Anselmino:1992vg}, not unlike electron Cooper pairs,
but this scenario is not the one under scrutiny here.

The purpose of this paper is to initiate the development of a
dynamical diquark {\em model}, thus turning the physical picture of
rapidly separating colored constituents into a formalism from which
quantifiable predictions for masses, selection rules for decay
channels, and branching fractions can be made under various
assignments of the relevant model parameters.

The first step in this direction is to describe how to express the
quantization of the $\delta$-$\bar \delta$ (or $\bar \theta$-$\delta$)
system in order to identify the appropriate combinations of quantum
numbers leading to the spectrum of mass eigenstates.  Already in the
initial presentation of the dynamical diquark
picture~\cite{Brodsky:2014xia}, such a configuration was described as
a color flux tube connecting the $\delta$-$\bar \delta$ pair, which
eventually absorbs all of the available kinetic energy in the process
until the pair comes relatively to rest.  The principal calculation of
Ref.~\cite{Brodsky:2014xia} supposed that the $Z_c(4430)^-$ resonance
appearing in $\Lambda_b \! \to \! (\pi^- \psi(2S) ) K^-$ results from
a $\delta$-$\bar \delta$ pair of known masses recoiling against the
$K^-$, the potential between them assumed to be of the classic
Coulomb-plus-linear Cornell type~\cite{Eichten:1978tg,Eichten:1979ms}.
The final separation of the $\delta$-$\bar \delta$ pair was calculated
to be 1.16~fm, comparable to (indeed, larger than) the expected
spatial extent of the $\psi(2S)$ wave function but much larger than
that of the $J/\psi$---providing a natural explanation why
$Z_c(4430)^-$ decays far more to $\psi(2S)$ than to
$J/\psi$~\cite{Chilikin:2014bkk}.

That the system should not undergo significant hadronization prior to
this moment is suggested by the Wentzel-Kramers-Brillouin (WKB)
approximation, which favors transitions when the configuration lies
near its classical turning point, since it gives an approximate wave
function
\begin{equation}
\psi(x) \simeq \frac{C}{\sqrt{p(x)}} e^{\pm \frac{i}{\hbar} \! \int \!
p(x) dx} \, ,
\end{equation}
where $p(x) \! = \! \sqrt{2\mu [E \! - \! V(x)]}$ is the classical
constituent relative momentum.  Such a configuration---two color
sources well separated and connected by strongly interacting field
configurations that may carry nontrivial quantum numbers---is
precisely the one used for heavy quarkonium hybrid studies,
particularly those performed on the lattice, an approach begun decades
ago~\cite{Griffiths:1983ah}.  In particular, the
{\it Born-Oppenheimer\/} (BO) {\it approximation}~\cite{Born:1927boa},
originally applied to atoms and molecules, proves instrumental for
studying systems containing heavy, slow-moving and light, fast-moving
degrees of freedom, which is the expectation for the $\delta$-$\bar
\delta$ state in its final moments prior to hadronization.

The dynamical diquark approach is rather different from those applied
to hidden-color multiquark states in the past, in that the state is
bound only by confinement until it can decay into hadrons, rather than
existing as a (quasi-)static configuration.  To provide context on
related works with different perspectives, Sec.~\ref{sec:Cousins}
briefly describes other directions for studying such states.

The BO approximation as applied to $\QQ$ exotics is reviewed briefly
in Sec.~\ref{sec:BO}.  We note immediately that its ground-state
multiplet, conventionally denoted by the BO potential $\Sigma^+_g$,
will be seen to coincide with the states one obtains from a
$\delta$-$\bar \delta$ Hamiltonian
approach~\cite{Maiani:2004vq,Maiani:2014aja}, and we enumerate and
develop a notation for these states even sooner, in
Sec.~\ref{sec:Ground}.  However, in the BO approach one need not
restrict to a single compact state with only contact interactions to
obtain this result; the same lowest multiplet occurs even if the state
has significant spatial extent.  For example, in the case of $\QQ$
hybrids, the $\Sigma^+_g$ multiplet simply represents conventional
quarkonium states without excited glue (as well as states in which the
glue content has all singlet quantum numbers).  In that case, the mass
gap between the lightest charmonium states ($\eta_c$, $J/\psi$) and
the lightest true hybrids, as calculated on the
lattice~\cite{Liu:2012ze}, is about 1.1--1.3~GeV\@.  However, since
the extended spatial structure of the $\Sigma^+_g$ states in the
dynamical diquark picture is expected to be similar to that of the
excited BO potential states, one may anticipate a smaller mass gap,
say in the several hundred MeV range.  They could be comparable to
typical radial excitation energies in charmonium [$m_{\psi(2S)} -
m_{J/\psi (1S)} \approx 600$~MeV], or even as small as typical orbital
excitations [$m_{\chi_{c} (1P)} - m_{J/\psi (1S)} \approx 400$~MeV].
If this possibility is realized, then the dynamical diquark-generated
phenomenological spectrum could be much richer than naively expected,
due to the intermingling of orbitally, radially, and BO-excited exotic
states.

By ``light'' quarks $q, q^\prime$ in this paper we mean only $u$ or
$d$, and by ``heavy'' quarks $Q$ we mean only $c$ or $b$.  However,
depending upon the circumstance, $s$ quarks may be considered light
(leading to a separate spectrum of, say, $c \bar c s \bar s$ exotics,
for which the $X(3915)$ state was proposed as the $J^{PC} \! = \!
0^{++}$ ground state, and exotics with the observed decay channel
$J/\psi \, \phi$ are natural candidates~\cite{Lebed:2016yvr}) or heavy
(for which one may seek interesting possible experimental signals of
their production, such as the small forward and backward enhancements
in the rate for $\gamma p \to \phi p$, which are consistent with a
pentaquark-like structure~\cite{Lebed:2015fpa}).  For the heavy
quarks, we have taken $Q$ and $\bar Q$ to have the same flavor ($b$ or
$c$), but exotic $B_c$-like hadrons, as well as doubly heavy $cc$,
$bb$, and $bc$ states, can also be studied within the BO approach,
although the treatment of the inversion quantum numbers $P$ and $C$
must be adjusted accordingly.

This paper is organized as follows.  Section~\ref{sec:Cousins}
presents a brief summary of alternate approaches studying multiquark
configurations interacting through color flux tubes or potentials, or
on the lattice.  In Sec.~\ref{sec:Ground}, we begin by enumerating the
states in the ground-state BO potential [$\Sigma^+_g(1L)$, $L \! = \!
0, 1, 2, \ldots$] in both $\delta$-$\bar \delta$ and heavy-quark spin
bases.  We then give a brief review of BO potentials as pertaining to
$\QQ q^\prime \! \bar q$ states in Sec.~\ref{sec:BO}, and in
Sec.~\ref{sec:BODiquark} identify the lowest expected potentials and
catalogue the corresponding spectra of $\delta$-$\bar \delta$ states
in Table~\ref{Table:states}\@.  In Sec.~\ref{sec:BOPenta} we carry out
this exercise for $\bar
\theta$-$\delta$ pentaquark states, with the results listed in
Table~\ref{Table:states2}\@.  Section~\ref{sec:Pheno} compares the
list of exotics with known quantum numbers to the spectra predicted in
the previous two sections, and then extends the analysis to discuss
constraints from heavy-quark spin symmetry and decay selection rules,
both exact and approximate.  In Sec.~\ref{sec:Concl} we summarize and
indicate future directions of research.

\section{Alternate Hidden-Color Multiquark Approaches}
\label{sec:Cousins}

By ``hidden color'' here, we mean states containing subunits carrying
nontrivial color charge, as opposed to only color-singlet (hadronic)
subunits.  These configurations were described many years ago in terms
color flux-tube structures connected in different topologies,
depending upon whether the shortest (energetically favored) tube
configurations connect $qq$ and $\bar q \bar q$ (diquark-like) pairs
or $q \bar q$ (meson-like) pairs.  The transition between the two was
called a ``flip-flop''~\cite{Miyazawa:1979vx,Oka:1984yx,Oka:1985vg}.
Numerical simulations have indeed found that such diquark-type
structures occur when the quarks are initially closer to each other
than to antiquarks, even when the relative distances are not
dramatically different~\cite{Alexandrou:2004ak,Okiharu:2004ve,
Cardoso:2011fq,Cardoso:2012uka,Bicudo:2017usw}.  If the coupling
between flux-tube configurations is weak, however,
Ref.~\cite{Carlson:1991zt} claims an absence of bound multiquark
states.

Not every simulation uncovers a diquark-antidiquark-like structure,
however. For example, Ref.~\cite{Ikeda:2013vwa} found no exotic $cc
\bar u\bar d$ state, but note that this state contains $cc$ rather than
$c\bar c$, and the diquarks are $(cc)(\bar u\bar d)$.  On the other
hand, Ref.~\cite{Prelovsek:2014swa} studies $\bar c c \bar d u$ but
does not find evidence for an exotic $Z_c^+$ state, but the authors
offer caveats that the basis of interpolating operators must be
carefully considered to include all coupled meson-me\-son states, and
that diquark-antidiquark operators having the same color structure
after Fierz rearrangement as such states.  Comments about the
incompleteness of the operator basis are also voiced in
Ref.~\cite{Guerrieri:2014nxa}.

Quark potential models with conventional two-body forces have also
been found to support four-quark structures, particularly in the
$Q\bar Q q\bar q$ case~\cite{Zouzou:1986qh}.  An extensive study of
possible multiquark structures~\cite{Lenz:1985jk} finds the presence
of hidden-color structures is quite feasible in the light-quark
sector.  Potentials based on confinement and instanton effects are
used to study $qq\bar q \bar q$ bound states in
Ref.~\cite{Beinker:1995qe}, but no good observed candidates are found.

However, quark models based solely on string-type confining
potentials~\cite{Vijande:2007ix} give more encouraging results for
$QQ\bar q \bar q$, although the authors suggest that bound $Q\bar Q q
\bar q$ bound states may require diquark substructure in order to
occur~\cite{Vijande:2009xx}. $QQ\bar q \bar q$ bound states also
appear in the more elaborate string-potential model calculations of
Ref.~\cite{Bicudo:2015bra}.

From this discussion, it should be clear that the status of
calculations of multiquark states is a delicate matter, depending upon
the precise modeling of the state and its interactions.  It should be
equally clear that the dynamical diquark picture does not obviously
appear to fit into any of these paradigms, especially as the system
remains in a state of rapid change until the moment it decays.  In
order to model such a system appropriately, we therefore seek to
describe it based not upon static structures, but upon symmetries:
hence the introduction of the Born-Oppenheimer approximation.  We
start with the ground-state band where the symmetries are trivial, and
then turn to the description of the excited states.

\section{Ground-State Band}
\label{sec:Ground}

In order to identify the $\delta$-$\bar \delta$ states of the
$\Sigma^+_g$ BO potential, we employ notation as close as possible to
that of Ref.~\cite{Maiani:2014aja}.  Starting with the 4 quark spins
and no orbital excitation, one may couple the angular momenta to
obtain a state of total constituent spin $S$ in several different
orders, but the most convenient for our purpose are $(\QQ) + (\qq)$
and $(qQ) + (\bar q \bar Q)$.  The first option uses eigenstates of
heavy-quark spin, while the second uses eigenstates of diquark spin.
Of course, all orders of coupling are connected by the relevant
recoupling coefficients, which in this case are $9j$ symbols:
\begin{eqnarray}
\lefteqn{\left< (s_q \, s_{\bar q}) s_\qq , (s_Q \, s_{\bar Q}) s_\QQ
, S \, \right| \left. (s_q \, s_Q) s_\de , (s_{\bar q} \, s_{\bar Q})
s_\bde , S \right> } & & \nonumber \\
& = & \left( [s_\qq] [s_\QQ] [s_\de] [s_\bde] \right)^{1/2}
\left\{ \begin{array}{ccc} s_q & s_{\bar q} & s_\qq \\
s_Q & s_{\bar Q} & s_\QQ \\ s_\de & s_\bde & S \end{array} \! \right\}
\, , \label{eq:9jTetra}
\end{eqnarray}
where $[s] \equiv 2s+1$ simply denotes the multiplicity of a spin-$s$
state.  Although we have here implicitly taken the light quarks $q,
\bar q$ to form a charge-conjugate pair, it is simple to generalize to
the case $q, {\bar q}^\prime$, where $q, q^\prime \in \{ u, d \}$,
which generates an $I \! = \! 0$ and three $I \! = \! 1$ states.  The
neutral ($I_3 \! = \! 0$) isosinglet and isotriplet eigenstates carry
definite $C$ eigenvalues, and in both cases the $C$-parity of these
states can be used to determine for all of the states the $G$-parity
eigenvalues $C (-1)^I$.

Although at this stage we still consider only $S$-wave states, let us
discuss the spatial-inversion parity eigenvalues $P$ and $C$ for
arbitrary $L$.  Using the usual reasoning applied to the corresponding
eigenvalues for conventional $\qq$ mesons, $P$ contains a factor
$(-1)^L$ from the inversion properties of orbital wave functions and a
$(-1)$ from the intrinsic parity of each $\qq$ pair (both light and
heavy).  Charge conjugation of the $\qq$ pairs is equivalent to a
combination of spatial inversion and the sign obtained from exchange
of the $q$, $\bar q$ spins, $(-1)^{s_\qq + 1}$.  One therefore obtains
the eigenvalues
\begin{equation} \label{eq:PCground}
P = (-1)^L \, , \ \ C = (-1)^{L + s_\qq + s_\QQ} \, .
\end{equation}
In particular, all $S$-wave tetraquarks have $P = +$, and the $(\qq),
(\QQ)$ basis is more convenient for identifying $C$ (and $G$)
eigenstates than the $\delta, \bar \delta$ basis.  In the $(\qq),
(\QQ)$ basis, and using the notation of Ref.~\cite{Maiani:2014aja},
one obtains the following states:
\begin{eqnarray}
J^{PC} = 0^{++}: & & X_0 \equiv \frac{1}{2} \left| 0_\qq , 0_\QQ
\right>_0 + \frac{\sqrt{3}}{2} \left| 1_\qq , 1_\QQ \right>_0 \, ,
\nonumber \\
& & X_0^\prime \equiv \frac{\sqrt{3}}{2} \left| 0_\qq , 0_\QQ
\right>_0 - \frac{1}{2} \left| 1_\qq , 1_\QQ \right>_0 \, , 
\nonumber \\
J^{PC} = 1^{++}: & & X_1 \equiv \left| 1_\qq , 1_\QQ \right>_1 \, ,
\nonumber \\
J^{PC} = 1^{+-}: & & Z \equiv \frac{1}{\sqrt 2} \left( \left| 1_\qq ,
0_\QQ \right>_1 \! - \left| 0_\qq , 1_\QQ \right>_1 \right) \, ,
\nonumber \\
& & Z^\prime \equiv \frac{1}{\sqrt 2} \left( \left| 1_\qq ,
0_\QQ \right>_1 \! + \left| 0_\qq , 1_\QQ \right>_1 \right) \, ,
\nonumber \\
J^{PC} = 2^{++}: & & X_2 \equiv \left| 1_\qq , 1_\QQ \right>_2 \, ,
\label{eq:SwaveQQ}
\end{eqnarray}
where outer subscripts indicate total component spin $S$.

Equivalently, using Eq.~(\ref{eq:9jTetra}), the states in the $\delta,
\bar \delta$ basis read:
\begin{eqnarray}
J^{PC} = 0^{++}: & & X_0 = \left| 0_\de , 0_\bde \right>_0 \, , \ \
X_0^\prime = \left| 1_\de , 1_\bde \right>_0 \, , \nonumber \\
J^{PC} = 1^{++}: & & X_1 = \frac{1}{\sqrt 2} \left( \left| 1_\de ,
0_\bde \right>_1 \! + \left| 1_\de , 0_\bde \right>_1 \right) \, ,
\nonumber \\
J^{PC} = 1^{+-}: & & Z = \frac{1}{\sqrt 2} \left( \left| 1_\de ,
0_\bde \right>_1 \! - \left| 0_\de , 1_\bde \right>_1 \right) \, ,
\nonumber \\
& & Z^\prime = \left| 1_\de , 1_\bde \right>_1 \, ,
\nonumber \\
J^{PC} = 2^{++}: & & X_2 = \left| 1_\de , 1_\bde \right>_2 \, .
\label{eq:Swavediquark}
\end{eqnarray}

Note that the pairs $X_0, X_0^\prime$ and $Z, Z^\prime$, carrying the
same $J^{PC}$, can certainly mix.  If one requires a basis of states
with definite values of heavy-quark spin, then the most convenient
combinations are:
\begin{eqnarray}
{\tilde X}_0 & \equiv & \left| 0_\qq , 0_\QQ \right>_0 =
+ \frac{1}{2} X_0 + \frac{\sqrt{3}}{2} X_0^\prime \, , \nonumber \\
{\tilde X}_0^\prime & \equiv & \left| 1_\qq , 1_\QQ \right>_0 =
+ \frac{\sqrt{3}}{2} X_0 - \frac{1}{2} X_0^\prime \, , \nonumber \\
{\tilde Z} & \equiv & \left| 1_\qq , 0_\QQ \right>_1 =
\frac{1}{\sqrt{2}} \left( Z^\prime \! + Z \right) \, , \nonumber \\
{\tilde Z}^\prime & \equiv & \left| 0_\qq , 1_\QQ \right>_1 =
\frac{1}{\sqrt{2}} \left( Z^\prime \! - Z \right) \, .
\label{eq:HQbasis}
\end{eqnarray}
Whenever the symbols $X_0$, $X^\prime_0$ ($Z$, $Z^\prime$) appear
below, it should be understood that ${\tilde X}_0$, ${\tilde
X}^\prime_0$ (${\tilde Z}$, ${\tilde Z}^\prime$) work equally well,
while the forms with tildes are specified if states of definite
$s_\QQ$ eigenvalues are preferred.

Turning next to the $L \! > \! 0$ states in the ground-state band, one
may use the usual rules of angular momentum addition to derive the
spectrum based upon the states $X_0$, $X^\prime_0$, $X_1$, $Z$,
$Z^\prime$, $X_2$ listed in Eq.~(\ref{eq:SwaveQQ}) or
(\ref{eq:Swavediquark}), by appending a subscript letter for the $L$
eigenvalue and a superscript number in parentheses for the total $J$
eigenvalue:
\begin{eqnarray}
\lefteqn{\underline{J^{PC} = L_{}^{(-1)^L \! , \, (-1)^L}}} & &
\nonumber \\
& & X_{0 \, L}^{(L)} , \ X_{0 \, L}^{\prime \, (L)} \, ,
\nonumber \\
\lefteqn{\underline{J^{PC} = (L-1, L, L+1)^{(-1)^L \! , \, (-1)^L}}}
& & \nonumber \\
& & X_{1 \, L}^{(L-1)} \! , \ X_{1 \, L}^{(L)} , \
X_{1 \, L}^{(L+1)} \! , \nonumber \\
\lefteqn{\underline{J^{PC} = (L-1, L, L+1)^{(-1)^L \! , \,
(-1)^{L+1}}}}  & & \nonumber \\
& & Z_L^{(L-1)} , \ \ Z_L^{(L)} , \ \ Z_L^{(L+1)} \! , \nonumber \\
& & Z_L^{\prime \, (L-1)} , \ Z_L^{\prime \, (L)} , \
Z_L^{\prime \, (L+1)} \! , \nonumber \\
\lefteqn{\underline{J^{PC} = (L-2, L-1, L, L+1, L+2)^{(-1)^L \! , \,
(-1)^L}}} & & \nonumber \\
& & X_{2 \, L}^{(L-2)} \! , \ X_{2 \, L}^{(L-1)} \! , \
X_{2 \,L}^{(L)} , \ X_{2 \, L}^{(L+1)} \! , \ X_{2 \, L}^{(L+2)} \, .
\label{eq:AllStates}
\end{eqnarray}
The $S$-wave states of Eq.~(\ref{eq:SwaveQQ}) or
(\ref{eq:Swavediquark}) are of course only those in
Eq.~(\ref{eq:AllStates}) with the largest $J$ value in each category:
$X_0 \! = \! X_{0 \, S}^{(0)}$, $X_0^\prime \! = \!  X_{0 \,
S}^{\prime \, (0)}$, $X_1 \! = \! X_{1 \, S}^{(1)}$, $Z \! = \!
Z_S^{(1)}$, $Z^\prime \! = \!  Z_S^{\prime \, (1)}$, $X_2 \!  = \!
X_{2 \, S}^{(2)}$.  For the $P$-wave states, one must also eliminate
the first two states in the final category: Explicitly, one obtains
the states $J^{PC} = 2 \times 1^{--}$ [$X_{0 \, P}^{(1)}, \, X_{0 \,
P}^{\prime \, (1)}$], $(0,1,2)^{--}$ [$X_{1 \, P}^{(0),(1),(2)}$], $2
\times (0,1,2)^{-+}$ [$Z_{P}^{(0),(1),(2)}, \, Z_{P}^{\prime \,
(0),(1),(2)}$], and $(1,2,3)^{--}$ [$X_{2 \, P}^{(1),(2),(3)}$].  In
this notation, the exhaustive list of $1^{--}$ states given in
Ref.~\cite{Maiani:2014aja} obtained by allowing all values of $L$
reads
\begin{eqnarray}
& & 
Y_1 \equiv X_{0 \, P}^{(1)} \, , \
Y_2 \equiv X_{1 \, P}^{(1)} \, , \
Y_3 \equiv X_{0 \, P}^{\prime \, (1)} \, , \nonumber \\ & &
Y_4 \equiv X_{2 \, P}^{(1)} \, , \
Y_5 \equiv X_{2 \, F}^{(1)} \, .
\end{eqnarray}
As a final illustration, the list of $D$-wave states reads $J^{PC} \!
= \! 2 \times 2^{++}$ [$X_{0 \, D}^{(2)}$, $X_{0 \, D}^{\prime \,
(2)}$], $(1,2,3)^{++}$ [$X_{1 \, D}^{(1),(2),(3)}$], $2 \times
(1,2,3)^{+-}$ [$Z_D^{(1),(2),(3)}$, $Z_D^{\prime \, (1),(2),(3)}$],
and $(0,1,2,3,4)^{++}$ [$X_{2 \, D}^{(0),(1),(2),(3),(4)}$].

Let us compare the number of conventional quarkonium states to the
number of tetraquark states listed above, including isospin.  For $L
\! = \! 0, 1, 2, 3, \ldots$, one counts $2,4,4,4,\ldots$ conventional
states [the usual $\eta, \psi \, ({\rm or} \ \Upsilon), h, \chi$
combinations] and $24,56,64,64,\ldots$ tet\-ra\-quark
states.\footnote{This counting for the $L \! = \! 0$ and $L \! = \! 1$
states was carried out in Ref.~\cite{Cleven:2015era}.}  Again, this
counting represents only the (radial) ground-state band (corresponding
to a principal quantum number $n \! = \! 1$), but it does count all
isospin states separately.  Not counting $I_3 \! = \! \pm 1$ charge
conjugates as distinct, the numbers reduce by 25\%, to
$18,42,48,48,\dots$.  While to date, 28 bosonic charmoniumlike exotics
have been observed (not counting charge conjugates), this number pales
in comparison to that for potential future discoveries, should even a
fraction of the predicted states actually exist.  To emphasize this
point, note that not even all of the $n \! = \! 1$ $D$-wave {\em
conventional\/} quarkonium states have yet been seen.

A well-known problem for diquark models is their tendency to produce
large numbers of unobserved states.  This overabundance occurs because
any $Qq$ or $\bar Q \bar q$ pair is considered suitable for forming a
diquark, regardless of its spin: The expectation in the light-quark
sector for nature to prefer a ``good'' (spin-0) diquark over a ``bad''
(spin-1) diquark~\cite{Jaffe:2004ph} is greatly reduced for
heavy-quark systems (since these two types of quasiparticle differ in
mass by a heavy-quark spin flip, which costs an energy proportional to
$\Lambda_{\rm QCD}^2/m_Q$), so that both types of diquark are expected
to be equally prevalent.  Furthermore, since isospin symmetry of
strong interactions implies that the replacement of $u \leftrightarrow
d$ quarks makes little change to the heavy diquarks, then in the
absence of significant isospin-dependent interactions between the
diquarks, one naturally expects tetraquarks formed of such diquarks to
appear in nearly degenerate $I \! = \! 0$ plus $I \! = \! 1$ quartets.

However, in the dynamical diquark picture as opposed to traditional
Hamiltonian-based diquark models, significant isospin-dependent
interactions may be quite natural due to the extended spatial size of
the state.  In the case of hadronic molecules, the long-distance
color-singlet attraction is expected to be dominated by single-pion
exchange since it is by far the lightest hadron, and in turn the pion
is light and carries nontrivial isospin due to the Nambu-Goldstone
(NG) theorem of chiral symmetry breaking.  Interestingly, a version of
the NG theorem exists even for colored particles (in the context of
{\it color-flavor locking\/}~\cite{Alford:1998mk}), so it is
reasonable to expect interactions with both color and isospin
dependence between the separated, colored diquarks.  The analysis of
Ref.~\cite{Cleven:2015era} argued in the case of hadronic molecules
that each of the two light-quark containing mesons contributes an
isospin Pauli matrix $\bm{\tau}_{(k)}$ to the interaction, and
$\bm{\tau}_{(1)} \cdot \bm{\tau}_{(2)} = -3,+1$ for $I \! = \!  0,1$,
meaning that the interaction is binding in one isospin channel and
repulsive in the other.  Therefore, one expects {\em only one\/} of
the $I \! = \! 0$ or $I \! = \! 1$ states for given angular momentum
quantum numbers to be bound, which greatly reduces the expected number
of tetraquark states, assuming a long-distance isospin-dependent
interaction between the colored diquarks.  Determining exactly which
of the states are bound of course requires a detailed model.

\section{Born-Oppenheimer Potentials}
\label{sec:BO}

The Born-Oppenheimer approximation amounts to a scale separation
between heavy, slowly changing degrees of freedom (hence effectively
acting as static sources) and light degrees of freedom (d.o.f.) that
rapidly and adiabatically adjust to the configuration of the heavy
ones.  The full wave function then factors into a part due to the
heavy sources, and a part described by {\it Born-Oppenheimer
potentials\/} that carry only the quantum numbers of the light d.o.f.\
but parametrically depend upon the configuration of the heavy sources
(hence the term ``potentials'').  In the original application to atoms
and molecules, these d.o.f.\ are of course the nuclei (mass $m_N$) and
electrons (mass $m_e$), respectively.  The scale separation, expressed
in powers of $m_e/m_N$, provides the necessary small parameter to
recast the BO approximation into the modern language of effective
field theories~\cite{Brambilla:2017uyf}.  In heavy quarkonium, the
$\QQ$ pair provides the static sources, while the light d.o.f.\ are
the gluon configuration (for hybrid mesons) or can also include
light-quark d.o.f.\ (for multiquark mesons)~\cite{Braaten:2014qka}.
The effective-field theory description arising from the BO
approximation for the hybrid case (where the expansion parameter
becomes $\Lambda_{\rm QCD}/m_Q$) was first considered in
Ref.~\cite{Berwein:2015vca}.

The configuration of the heavy d.o.f.\ is described both by the
relative separations of the heavy components and by its symmetry.  In
the $\QQ$ system with a relative separation $r$ and a unit vector
$\hat {\bm{r}}$ pointing from $\bar Q$ to $Q$, the BO potential
depends only upon $r$, and the potentials are labeled by the
irreducible representations of the group $D_{\infty h}$, which
describes the symmetries of a cylinder with axis $\hat {\bm{r}}$.  The
conventional nomenclature~\cite{Landau:1977} for these representations
uses the quantum numbers $\Gamma \equiv \Lambda^\epsilon_\eta$, all of
which refer to the $D_{\infty h}$ symmetry, as we now describe.

The basic angular momenta of the system are the total $\bm{J}_{\rm
light}$ of the light d.o.f., the orbital angular momentum $\bm{L}_\QQ$
of the heavy d.o.f., and spin $\bm{s}_\QQ$ of the $\QQ$ pair.  Due to
heavy-quark symmetry, $s_\QQ$ is a good quantum number of the full
state, but $\bm{J}_{\rm light}$ and $\bm{L}_\QQ$ cannot be
independently determined, although the Casimirs $J_{\rm light}$ and
$L_\QQ$ can be simultaneously specified.  In this definition, the
light-quark spin $s_\qq$ (in the case of multiquark hadrons) is
incorporated into $\bm{J}_{\rm light}$.  One then defines the total
orbital angular momentum as
\begin{equation} \label{eq:Ldef}
\bm{L} \equiv \bm{L}_\QQ + \bm{J}_{\rm light} \, ,
\end{equation}
and finally, from coupling $L$ and $s_\QQ$, one obtains the total
angular momentum quantum numbers $J, J_z$ of the state.  Since $\hat
{\bm{r}} \! \cdot \! \bm{L}_\QQ = 0$, the axial angular momentum $\hat
{\bm{r}} \! \cdot \! \bm{J}_{\rm light} \! = \hat {\bm{r}} \!
\cdot \! \bm{L}$ for the light d.o.f.\ provides a good quantum number
for the system, its eigenvalues denoted by $\lambda = 0, \pm 1, \pm 2,
\ldots$.  Since the physical system is invariant under a reflection
through any plane containing $\hat {\bm{r}}$ (under which $\lambda \!
\to \! -\lambda$), its energy eigenvalues cannot depend upon the sign
of $\lambda$, and from this fact one defines the first of the BO
quantum numbers, $\Lambda \equiv |\lambda|$.  Potentials with the
eigenvalues $\Lambda=0,1,2,\ldots$ are denoted by
$\Sigma,\Pi,\Delta,\ldots$, in analogy to the labels $S,P,D,\ldots$
for the quantum numbers $L=0,1,2,\ldots$.  From Eq.~(\ref{eq:Ldef})
and $\hat {\bm{r}} \! \cdot \! \bm{L}_\QQ = 0$, one immediately notes
the constraint
\begin{equation} \label{eq:LambdaMax}
L \ge |\hat {\bm{r}} \cdot \bm{L}| = |\hat {\bm{r}} \cdot
\bm{J}_{\rm light}| = \Lambda \, .
\end{equation}

The light d.o.f.\ also possess two reflection symmetries.  The first
is obtained by a reflection through the midpoint of the $\QQ$ pair.
Since this inversion exchanges the orientation of the light d.o.f.\
not just with respect to a coordinate origin but also with respect to
$Q$ and $\bar Q$, it is given not just by the parity operator $P_{\rm
light}$, but in fact by the combination $(CP)_{\rm light}$.  Its
possible eigenvalues $\eta = +1 , -1$, denoted by $g , u$,
respectively, provide the second BO quantum number.

The system also possesses, as mentioned above, a symmetry under
reflection $R_{\rm light}$ of the light d.o.f.\ through any plane
containing the $\QQ$ axis.  In particular, the $\Lambda \! = \! 0$
($\Sigma$) representations can be distinguished by their behavior
under $R_{\rm light}$, with its $\pm 1$ eigenvalue denoted by
$\epsilon$, the third BO quantum number.  But the $\Lambda > 0$
configurations $\left| \lambda, \eta; \bm{r} \right>$ can also be
combined into eigenstates of $R_{\rm light}$ with eigenvalue
$\epsilon$: Noting that the light d.o.f.\ spatial-inversion parity
operator $P_{\rm light}$ is simply given by $R_{\rm light}$ multiplied
by a rotation by $\pi$ radians about an axis normal to the plane
defining $R_{\rm light}$, one sees for arbitrary $\lambda$ that
$R_{\rm light} \left| \lambda, \eta; \bm{r} \right> = (-1)^\lambda
\zeta \left| -\lambda, \eta; \bm{r} \right>$, where $\zeta$ is the
intrinsic parity of the light d.o.f\@.  The eigenstate of $R_{\rm
  light}$ with eigenvalue $\epsilon$ for $\Lambda > 0$ is then
constructed as
\begin{equation}
  \left| \Lambda, \eta, \epsilon; \bm{r} \right> \equiv
  \frac{1}{\sqrt{2}} \left[ \left| \Lambda, \eta; \bm{r} \right>
    + \epsilon \, (-1)^\Lambda \zeta \left| -\Lambda, \eta; \bm{r}
  \right> \right] \, ,
\end{equation}
and the eigenvalue of $P_{\rm light}$ is deduced to be $\epsilon \,
(-1)^\Lambda$.

With the quantum numbers $\Gamma$ in hand, one then solves the
Schr\"{o}dinger equation of the $\QQ$ pair in the BO potential
$V_\Gamma(r)$, which produces eigenvalues and eigenfunctions labeled
by a principal quantum number $n$.  The full physical states are then
completely specified by the kets
\begin{equation}
\left| n, L, s_\QQ, J m_J; \Lambda, \eta, \epsilon \right> \, ,
\end{equation}
with $J_{\rm light}$ and $L_\QQ$ eigenvalues implicit.  In the
multiquark case, the light-quark spin quantum number $s_\qq$ is also
implicit, providing in the notation of Ref.~\cite{Braaten:2014qka} a
contribution to $\bm{J}_{\rm light}$.

The overall discrete quantum numbers for the physical state depend
upon both the heavy and light d.o.f\@.  Those for the heavy d.o.f.\
$\QQ$ are obtained exactly as for ordinary mesons, while those for the
light d.o.f.\ depend upon whether a $\qq$ pair is present, which
contributes an extra factor $(-1)$ to $P$ and $(-1)^{s_\qq}$ to $C$.
In particular, for hybrids,
\begin{eqnarray}
P & = & \; \; \epsilon \, (-1)^{\Lambda + L + 1} \, ,
\label{eq:Pval} \\
C & = & \eta \epsilon \, (-1)^{\Lambda + L + s_\QQ} \, ,
\label{eq:Cval}
\end{eqnarray}
while for tetraquarks,
\begin{eqnarray}
P & = & \; \; \epsilon \, (-1)^{\Lambda + L} \, , \label{eq:Pval2} \\
C & = & \eta \epsilon \, (-1)^{\Lambda + L + s_\qq + s_\QQ} \,
. \label{eq:Cval2}
\end{eqnarray}
The $C$ eigenvalue, as before, refers to that of the neutral state of
an isospin multiplet; $G$ parity is then given by $G = C(-1)^I$.
Significantly, the expressions Eqs.~(\ref{eq:Pval2})--(\ref{eq:Cval2})
differ from those in Ref.~\cite{Braaten:2014qka}, which are the same
for both hybrids and tetraquarks.  Even though the light-quark pair
has its spin angular momentum $s_\qq$ folded into the total $J_{\rm
light}$ in Ref.~\cite{Braaten:2014qka}, including its distinct
dependence in $P$ and $C$ is necessary to reflect the differing
symmetry of the wave functions, especially for differing values of
$s_\qq$, which already suggests difficulties for the choice of
including $s_\qq$ in $J_{\rm light}$.  In particular, one expects
states that are identical except for a relative spin flip of the light
quarks, $s_\qq \! = \! 0 \!
\leftrightarrow \! s_\qq \! = \! 1$, to belong to the same BO potential
(fixed $\Gamma \! = \! \Lambda^\epsilon_\eta$), but also to have
opposite $C$ eigenvalues.  This effect is particularly evident in the
ground-state band $\Sigma_g^+$ ($\Lambda \! = \! 0$, $\epsilon \! = \!
\eta \! = \!  +1$), where one may use simple quark-model reasoning as
in Eq.~(\ref{eq:PCground}).  As an explicit example, in the case of
$b\bar b c\bar c$ tetraquarks, for which $m_b \! \gg \! m_c \! \gg \!
\Lambda_{\rm QCD}$, one expects flipping the spin of $\bar c$ relative
to that of $c$ (using the definition in which $s_\qq$ is a part
of $J_{\rm light}$) to affect the value of $J_{\rm light}$ and hence
$\Lambda$, which would spread these two configurations over different
BO potentials.  But the energy cost of this spin flip is small,
$O(\Lambda_{\rm QCD}^2/m_c)$, suggesting that the BO potentials in the
two configurations are the same.

We therefore adopt a more traditional definition of quantum numbers
for BO potentials~\cite{Landau:1977}: The angular momentum
$\bm{J}_{\rm light}$ in Eqs.~(\ref{eq:Ldef}) and (\ref{eq:LambdaMax})
is understood to exclude intrinsic light-quark spin $\bm{s}_\qq$, and
the BO potential notation becomes $\Gamma \equiv {}^{2s_\qq+1} \!
\Lambda^\epsilon_\eta$, the new superscript indicating the
multiplicity of $(s_\qq)_z$ eigenstates.  From the above example, one
also expects the configurations ${}^3 \!
\Lambda^\epsilon_\eta (nP)$ and ${}^1 \!  \Lambda^\epsilon_\eta (nP)$
to lie fairly close in energy, ignoring possible light-quark
spin-dependent interactions such as those correlated with isospin.  We
therefore suppress the $2s_\qq \! + \! 1$ superscript whenever
possible.

Also of interest is the possibility of $\Lambda$-{\it
doubling\/}~\cite{Landau:1977}, which occurs when two BO potentials
produce the same spectrum of states, and therefore can mix.  For given
eigenvalues of $L$ and $\Lambda$ satisfying $L \! \ge \! 1$ and $L \!
> \! \Lambda$, the states obtained from the BO potentials
$\Lambda^\epsilon_\eta (nL)$ and $(\Lambda + 1)^{-\epsilon}_\eta (nL)$
[{\it e.g.}, $\Sigma^-_u (1P)$ and $\Pi^+_u (1P)$] produce the same
spectrum, and potentially can mix.  The naive degeneracy between two
BO potentials of opposite parity, $\Lambda^\pm_\eta (nL)$ [{\it e.g.},
$\Pi^\pm_u (1P)$], is thereby lifted.  This effect for hybrids was
first discussed in Ref.~\cite{Berwein:2015vca}.

\section{Diquark-Antidiquark BO Potentials}
\label{sec:BODiquark}

The configuration of the tetraquark state in the dynamical diquark
picture is essentially the same as for hybrid heavy-quark mesons: a
spatially extended colored field connecting a heavy color-{\bf 3}
($\bar \delta$) source and a heavy color-$\bar{\bf 3}$ source
($\delta$).  The sources themselves differ in the two cases: $Q, \bar
Q$ carry spin $\frac 1 2$ and isospin 0, and are essentially
pointlike, while $\delta$, $\bar \delta$ carry spin 0 or 1 and isospin
$\frac 1 2$, and are expected to be compact due to the presence of the
heavy quark but still be of finite spatial extent ($\lesssim$~0.5~fm
for charm~\cite{Brodsky:2014xia}).

The static potentials for $\QQ$ pairs were first calculated on the
lattice some time ago, with the first high-quality results presented
in Refs.~\cite{Juge:1997nc,Juge:1999ie,Juge:2002br}, while the first
unquenched simulations were carried out in Ref.~\cite{Bali:2000vr}.  A
summary of the important landmarks in lattice simulations relevant to
heavy-quark hybrids is presented in Ref.~\cite{Berwein:2015vca} (see
also~\cite{Lebed:2017xih}).  Simulations representing the state of the
art for $c\bar c$ hybrid mesons are presented by the Hadron Spectrum
Collaboration in Ref.~\cite{Liu:2012ze}.  The essential result
relevant to the present analysis is that all authors agree the lowest
BO potentials are determined to be the ground state $\Sigma^+_g$,
followed by $\Pi_u$ and $\Sigma^-_u$.  The mixing of $\Pi^+_u(1P)$ and
$\Sigma^-_u(1P)$ states has been noted in the previous section; but
additionally, the $\Pi_u$ and $\Sigma^-_u$ BO potentials are seen to
become degenerate in the $r \! \to \! 0$ limit, giving a single
color-adjoint source configuration as $r \! \to \! 0$ (in this case,
with $J^{PC} = 1^{+-}$) called a {\it gluelump}.

We therefore suppose that the lowest BO potentials producing
tetraquarks in the dynamical diquark picture are the ground-state
potentials $\Sigma^+_g$, whose $S$-, $P$-, and $D$-wave states have
already been enumerated in Sec.~\ref{sec:Ground}, followed by
$\Pi_u^+(1P)$ mixed with $\Sigma_u^-(1P)$, $\Pi_u^-(1P)$, $\Sigma_u^-
(1S)$, and $\Pi_u^+(1D)$.  This ordering follows the results of the
lattice simulations of Ref.~\cite{Liu:2012ze}, with the states
identified as originating within specific BO potentials in
Refs.~\cite{Braaten:2014qka,Berwein:2015vca}.

Before listing the spectra associated with these BO potentials, let us
make one final modification to the notation $(X_0, X^\prime_0, X_1, Z,
Z^\prime, X_2)^{(J)}_L$ introduced for the $\Sigma^+_g$ states.
Noting from Eqs.~(\ref{eq:Pval2})--(\ref{eq:Cval2}) that the $P,C$
eigenvalues for nontrivial BO potentials $\Lambda^\epsilon_\eta$
differ from those of $\Sigma^+_g$ only by
\begin{equation}
\rho \! \equiv \! \epsilon \, (-1)^\Lambda, \
\kappa \! \equiv \! \eta \epsilon \, (-1)^\Lambda \! =
\!  \eta \rho \, ,
\end{equation}
we adopt the final notation $(X_0, X^\prime_0, X_1, Z, Z^\prime,
X_2)^{(J) \rho \kappa}_L$ for the tetraquark states.  That is,
Eqs.~(\ref{eq:Pval2})--(\ref{eq:Cval2}) are replaced by
\begin{equation} \label{eq:FinalPC}
P = \rho \, (-1)^L , \ C = \kappa \, (-1)^{L + s_\qq +
s_\QQ} \, ,
\end{equation}
which identifies $\rho, \kappa$ as the ``intrinsic'' $P, C$
eigenvalues of each particular BO potential.  In this notation, one
appends a superscript $\rho \kappa \! = \! ++$ to all the states
obtained from $\Sigma^+_g$, and taking $\epsilon \!
\to \!  -\epsilon$ or $\Lambda \! \to \! \Lambda \pm 1$ changes both
$\rho \! \to \! -\rho$ and $\kappa \! \to \! -\kappa$, while taking
$\eta \! \to \! -\eta$ changes $\kappa \! \to \! -\kappa$ alone.
Taking $\epsilon \! \to \! -\epsilon$ or $\Lambda \! \to \! \Lambda
\pm 1$ or $L \! \to \! L \! \pm \! 1$ changes both $P \! \to \! -P$
and $C \! \to \! -C$ for each state in the BO potential, while taking
$\eta \! \to \! -\eta$ changes only $C \! \to \! -C$ for each state.

To see that this notation is easily interpreted, let us consider one
specific example: $X_{2 \, F}^{(4) \, -+}$.  Here, the total component
spin $S \! = \! 2$ state $X_2$ is defined in Eq.~(\ref{eq:SwaveQQ})
with $s_\qq \! = \!  1$, $s_\QQ \! = \! 1$; in addition,
$L \! = \! 3$, $J \! = \! 4$, $\rho \! = \!  \epsilon \, (-1)^\Lambda
\!  = \! -1$ so that $P \! = \! +$, and $\kappa \!  = \! \eta \rho \! =
\!  +1$ so that $\eta \! = \! -1$ and $C \! = \! -$: $X_{2 \, F}^{(4)
\, -+}$ is a $J^{PC} \! = \! 4^{+-}$ state.  The only ambiguity lies
in the combination $\rho \!  = \!  \epsilon \, (-1)^\Lambda \! = \!
-1$; since $\Lambda \! \le \! L \!  = \!  3$, the BO potentials
$[\Sigma^-_u, \Pi^+_u, \Delta^-_u, \Phi^+_u](nF)$ can contribute.

In Table~\ref{Table:states} we list the lowest multiplets of
tetraquark states expected in the dynamical diquark picture, both by
$J^{PC}$ eigenvalues and the BO potential from which they emerge.
States that for $\qq$ mesons have exotic $J^{PC}$ quantum numbers
(specifically, $0^{--}$ and the series $0^{+-}, 1^{-+}, 2^{+-},
\ldots$) are indicated with boldface.  We also use the $(\qq),
(\QQ)$ basis [Eq.~(\ref{eq:HQbasis})] in order to facilitate
comparison in Sec.~\ref{sec:Pheno} with the expectations of
heavy-quark spin symmetry.
\begin{table*}
  \caption{Quantum numbers of the lowest tetraquark states expected in
  the dynamical diquark picture.  For each of the expected lowest
  Born-Oppenheimer potentials, the full multiplet is presented, using
  both the state notation developed in this work and the corresponding
  $J^{PC}$ eigenvalues.  States with $J^{PC}$ not allowed for
  conventional $\qq$ mesons are indicated in boldface.}
\label{Table:states}
\begin{tabular}{c|c|c|c}
  \hline\hline
  BO potential & \multicolumn{3}{c}{State notation} \\
  & \multicolumn{3}{c}{State $J^{PC}$} \\
  \hline
  $\Sigma^+_g(1S)$ & $ \tilde X_{0 \, S}^{(0)++}$ &
  $\tilde Z_S^{(1) ++}$, $\tilde Z_S^{\prime \, (1) ++}$ &
  $\tilde X_{0 \, S}^{\prime \, (0) ++}$, $X_{1 \, S}^{(1) ++}$,
  $X_{2 \, S}^{(2) ++}$ \\
  & $0^{++}$ & $2 \times 1^{+-}$ & $[0,1,2]^{++}$ \\
  \hline
  $\Sigma^+_g(1P)$ & $\tilde X_{0 \, P}^{(1) ++}$ &
  $[\tilde Z_P^{(0), \bm{(1)}, (2)}]^{++}$,
  $[\tilde Z_P^{\prime \, (0), \bm{(1)}, (2)}]^{++}$ &
  $\tilde X_{0 \, P}^{\prime \, (1) ++}$, \
  $[X_{1 \, P}^{\bm{(0)}, (1), (2)}]^{++}$, \
  $[X_{2 \, P}^{(1), (2), (3)}]^{++}$ \\
  & $1^{--}$ & $2 \times (0,\bm{1},2)^{-+}$ &
  $[ 1, \ (\bm{0},1,2), \ (1,2,3) ]^{--}$ \\
  \hline
  $\Sigma^+_g(1D)$ & $\tilde X_{0 \, D}^{(2)++}$ &
  $[\tilde Z_D^{(1),\bm{(2)},(3)}]^{++}$,
  $[\tilde Z_D^{\prime \, (1),\bm{(2)},(3)}]^{++}$ &
  $\tilde X_{0 \, D}^{\prime \,  (2)++}$, \ 
  $[X_{1 \, D}^{(1),(2),(3)}]^{++}$, \
  $[X_{2 \, D}^{(0),(1),(2),(3),(4)}]^{++}$ \\
  & $2^{++}$ & $2 \times (1,\bm{2},3)^{+-}$
  & $[ 2, \ (1,2,3), \ (0,1,2,3,4)]^{++}$ \\
  \hline
  $\Pi^+_u(1P)$ \& & $\tilde X_{0 \, P}^{(1) -+}$ &
  $[\tilde Z_P^{(0), (1), (2)}]^{-+}$,
  $[\tilde Z_P^{\prime \, (0), (1), (2)}]^{-+}$ &
  $\tilde X_{0 \, P}^{\prime \, (1) -+}$, \
  $[X_{1 \, P}^{\bm{(0)}, (1), \bm{(2)}}]^{-+}$, \
  $[X_{2 \, P}^{(1), \bm{(2)}, (3)}]^{-+}$ \\
  $\Sigma^-_u(1P)$ & $1^{+-}$ & $2 \times (0,1,2)^{++}$ &
  $[ 1, \ (\bm{0},1,\bm{2}), \ (1,\bm{2},3) ]^{+-}$ \\
  \hline
  $\Pi^-_u(1P)$ & $\tilde X_{0 \, P}^{\bm{(1)} +-}$ &
  $[\tilde Z_P^{\bm{(0)}, (1), (2)}]^{+-}$,
  $[\tilde Z_P^{\prime \, \bm{(0)}, (1), (2)}]^{+-}$ &
  $\tilde X_{0 \, P}^{\prime \, \bm{(1)} +-}$, \
  $[X_{1 \, P}^{(0), \bm{(1)}, (2)}]^{+-}$, \
  $[X_{2 \, P}^{\bm{(1)}, (2), \bm{(3)}}]^{+-}$ \\
  & $\bm{1}^{-+}$ & $2 \times (\bm{0},1,2)^{--}$ &
  $[ \bm{1}, \ (0,\bm{1},2), \ (\bm{1},2,\bm{3}) ]^{-+}$ \\
  \hline
  $\Sigma^-_u(1S)$ & $ \tilde X_{0 \, S}^{(0)-+}$ &
  $\tilde Z_S^{(1) -+}$, $\tilde Z_S^{\prime \, (1) -+}$ &
  $\tilde X_{0 \, S}^{\prime \, (0) -+}$, $X_{1 \, S}^{\bm{(1)} -+}$,
  $X_{2 \, S}^{(2) -+}$ \\
  & $0^{-+}$ & $2 \times 1^{--}$ & $[0,\bm{1},2]^{-+}$ \\
  \hline
  $\Pi^+_u(1D)$ & $\tilde X_{0 \, D}^{(2)-+}$ &
  $[\tilde Z_D^{(1),(2),(3)}]^{-+}$,
  $[\tilde Z_D^{\prime \, (1),(2),(3)}]^{-+}$ &
  $\tilde X_{0 \, D}^{\prime \, (2)-+}$, \ 
  $[X_{1 \, D}^{\bm{(1)},(2),\bm{(3)}}]^{-+}$, \
  $[X_{2 \, D}^{(0),\bm{(1)},(2),\bm{(3)},(4)}]^{-+}$ \\
  & $2^{-+}$ & $2 \times (1,2,3)^{--}$
  & $[ 2, \ (\bm{1},2,\bm{3}), \ (0,\bm{1},2,\bm{3},4)]^{-+}$ \\
\hline\hline
\end{tabular}
\end{table*}

\section{Pentaquark BO Potentials}
\label{sec:BOPenta}

The central difference between diquark-antidiquark ($\delta$-$\bar
\delta$) and triquark-diquark ($\bar \theta$-$\delta$) BO potentials
is that the latter case is analogous to heteronuclear diatomic
molecules: The $\bar \theta$ and $\delta$ components are in no sense
the same, so that the reflection symmetry leading to the $(CP)_{\rm
light}$ quantum number $\eta$ is lost.  In addition, the
(anti)triquark $\bar \theta$ is formed of a {\em light\/} diquark
$\delta^\prime$ in a color-$\bar {\bf 3}$ bound to a heavy $\bar Q$ to
form an overall color-{\bf 3}.  For the purpose of this work, we limit
to the case of a $\delta^\prime \! = \! (ud)$ diquark in a spin-0,
isospin-0 configuration (a ``good'' diquark~\cite{Jaffe:2004ph}), such
as those naturally appearing in $\Lambda_Q$ baryons, $Q = s, c, b$.
Indeed, the pentaquark candidates $P_c(4380)$, $P_c(4450)$ were
observed in $\Lambda_b$ decays~\cite{Aaij:2015tga}, a fact used in the
construction of the diquark-triquark picture~\cite{Lebed:2015tna}.
Assuming (as for the diquark $\delta \! = \! Qq$) no internal orbital
angular momentum, the antitriquark $\bar \theta \equiv [\bar Q (ud)]$
carries the unique quantum numbers $s_{\bar \theta}^{P_{\bar
\theta}} = {\frac 1 2}^-$.  The intrinsic parity of the $\QQ qud$
pentaquark state is $-1$, due to the presence of the single antiquark
$\bar Q$; its isospin $I \! = \! \frac 1 2$ is determined entirely by
the light quark $q$ in $\delta$.

As noted in Ref.~\cite{Lebed:2015tna}, the Pauli exclusion principle
must be taken into account if $\delta^\prime$ contains identical
quarks (in which case it would cease to be a ``good'' diquark).  But
even if $\delta^\prime$ {\em is\/} a good diquark, then the light
quark in $\delta$ is identical to one of those in $\delta^\prime$, and
possible constraints on the overall state due to antisymmetrization
between these quarks must be considered.  Inasmuch as $\delta^\prime$
(as a part of the antitriquark $\bar \theta$) and $\delta$ are
expected to achieve substantial spatial separation in the dynamical
picture, the effect of antisymmetrization on matrix elements of
observables should be significantly muted.

The construction of the lowest pentaquark states uses the same
principles as used for the tetraquark states in
Secs.~\ref{sec:Ground}--\ref{sec:BODiquark}, so we present explicitly
in this section only the most important intermediate results.  Since
heavy-quark spin symmetry remains of interest in this system, we begin
by exhibiting the relation between the $\bar \theta$-$\delta$ basis
and the $(q \delta^\prime)(\QQ)$ basis, in which the light quark $q$
in $\delta$ is instead coupled with $\delta^\prime$ to form an
all-light baryonic system $B \equiv (q \delta^\prime)$.  Analogous to
Eq.~(\ref{eq:9jTetra}), it reads
\begin{eqnarray}
\lefteqn{\left< (s_q \, s_\dep) s_\B , (s_Q \, s_{\bar Q}) s_\QQ , S
\, \right| \left.
(s_q \, s_Q) s_\de , (s_\dep \, s_{\bar Q}) s_\bt , S \right> }
& & \nonumber \\
& = & \left( [s_\B] [s_\QQ] [s_\de] [s_\bt] \right)^{1/2}
\left\{ \begin{array}{ccc} s_q & s_\dep & s_\B \\
s_Q & s_{\bar Q} & s_\QQ \\ s_\de & s_\bt & S \end{array} \! \right\}
\, . \label{eq:9jPenta}
\end{eqnarray}
At this point, the diquark $\delta^\prime$ spin has not yet been fixed
to 0\@.  Making this restriction, however, one finds only 3 basis
states:
\begin{eqnarray}
J^{PC} = {\frac 1 2}^-: & &
P_{\frac 1 2} \equiv \left| \hf_\bt , 0_\de \right>_{\frac 1 2} , \ \
P^\prime_{\frac 1 2} \equiv \left| \hf_\bt , 1_\de \right>_{\frac 1 2}
, \nonumber \\
J^{PC} = {\frac 3 2}^-: & &
P_{\frac 3 2} \equiv \left| \hf_\bt , 1_\de \right>_{\frac 3 2} .
\label{eq:SwaveDiTri}
\end{eqnarray}
The corresponding list that includes both these states and also allows
$s_\dep \! = \! 1$ (giving 6 additional states) appears in
Ref.~\cite{Zhu:2015bba}, albeit using a different notation.

In terms of Eq.~(\ref{eq:SwaveDiTri}) and using
Eq.~(\ref{eq:9jPenta}), the states of definite heavy-quark spin can
then be written:
\begin{eqnarray}
{\tilde P}_{\frac 1 2} & \equiv & \left| \hf_B , 0_\QQ
\right>_{\frac 1 2}  = -\frac 1 2 P_{\frac 1 2} + \frac{\sqrt{3}}{2}
P^\prime_{\frac 1 2} \, , \nonumber \\
{\tilde P}^\prime_{\frac 1 2} & \equiv & \left| \hf_B , 1_\QQ
\right>_{\frac 1 2}  = +\frac{\sqrt{3}}{2} P_{\frac 1 2} + \frac 1 2
P^\prime_{\frac 1 2} \, , \nonumber \\
P_{\frac 3 2} & = & \left| \hf_B , 1_\QQ \right>_{\frac 3 2} \, .
\label{eq:PentaHQS}
\end{eqnarray}

The generalization of these states to $L \! > \! 0$, analogous to
Eq.~(\ref{eq:AllStates}), reads
\begin{eqnarray}
\lefteqn{\underline{J^P = (L -\hf , \, L +\hf)^{(-1)^{L+1}}}} & &
\nonumber \\
& & P_{\frac 1 2 \, L}^{(L-\frac 1 2)} \! , \
P_{\frac 1 2 \, L}^{\prime \, (L-\frac 1 2)} \! ; \
P_{\frac 1 2 \, L}^{(L+\frac 1 2)} \! ,
P_{\frac 1 2 \, L}^{\prime \, (L+\frac 1 2)} \! , \\
\lefteqn{\underline{J^P = (L -\h3 , \, L -\hf , \, L +\hf , \, L +\h3)
^{(-1)^{L+1}}}} & & \nonumber \\
& & P_{\frac 3 2 \, L}^{(L-\frac 3 2)} \! , \
P_{\frac 3 2 \, L}^{(L-\frac 1 2)} \! , \
P_{\frac 3 2 \, L}^{(L+\frac 1 2)} \! ,
P_{\frac 3 2 \, L}^{(L+\frac 3 2)} \! .
\label{eq:AllStates2}
\end{eqnarray}
Of course, any states in this list with $J$ disallowed by the triangle
rule $|L \!  - \! S| \! \le J \! \le \! L \! + \! S$ are excluded.

Since the $\bar \theta$-$\delta$ states are not eigenstates of
$(CP)_{\rm light}$ and hence lack $\eta$ (and consequently $C$)
eigenvalues, their nontrivial BO potentials are simply labeled by
$\Lambda^\epsilon$, and their states carry the parity eigenvalues
\begin{equation} \label{eq:Ppenta}
P = \epsilon \, (-1)^{\Lambda + L + 1} \equiv \rho \, (-1)^{L+1} \, .
\end{equation}
The final addition to the notation of Eq.~(\ref{eq:AllStates2}) is to
append the superscript $\rho$ defined in Eq.~(\ref{eq:Ppenta}) to the
state symbol.  The analogue to Table~\ref{Table:states} for $\bar
\theta$-$\delta$ states with $s_\dep \! = \! 0$, representing the
lowest expected pentaquark states in the triquark-diquark picture, is
presented as Table~\ref{Table:states2}.  Again, the notation of
Eq.~(\ref{eq:PentaHQS}) is employed, to enable comparisons with
expectations from heavy-quark spin symmetry.  A state such as, {\it
e.g.}, $\tilde P^{\prime \, ( \frac 3 2 ) +}_{\frac 1 2 \, D}$ means
$s_\B \! = \! \frac 1 2$, $s_\QQ \! = \! 1$, $S \! = \!  \frac 1 2$,
$L \! = \! 2$, $J \! = \! \frac 3 2$, $\rho \! = \! +$, and $P \! = \!
\rho \, (-1)^{L+1} \! = \! -$: As indicated in Table~\ref{Table:states2},
it has $J^P \! = \! {\frac 3 2}^-$.
\begin{table*}
  \caption{Quantum numbers of the lowest pentaquark states expected in
    the dynamical triquark-diquark picture.  For each of the expected
    lowest Born-Oppenheimer potentials, the full multiplet is
    presented, using both the state notation developed in this work
    and the corresponding $J^P$ eigenvalues.}
\label{Table:states2}
\setlength{\extrarowheight}{1.5ex}
\begin{tabular}{c|c|c}
  \hline\hline
  BO potential & \multicolumn{2}{c}{State notation} \\
  & \multicolumn{2}{c}{State $J^P$} \\
  \hline
  $\Sigma^+ (1S)$ & $\tilde P_{\frac 1 2 \, S}^{( \frac 1 2 )+}$,
  $\tilde P_{\frac 1 2 \, S}^{\prime \, ( \frac 1 2 )+}$ &
  $P_{\frac 3 2 \, S}^{( \frac 3 2 ) +}$ \\
  & $2 \times {\frac 1 2}^-_{\vphantom\dagger}$ & ${\frac 3 2}^-$ \\ 
  \hline
  $\Sigma^+ (1P)$ & $\Big[ \tilde P_{\frac 1 2 \, P}^{( \frac 1 2 ),
  (\frac 3 2 )} \Big]^+, \ \Big[ \tilde P_{\frac 1 2 \, P}^{\prime
  \, ( \frac 1 2 ), (\frac 3 2)} \Big]^+$ &
  $\Big[ P_{\frac 3 2 \, P}^{( \frac 1 2 ), ( \frac 3 2 ),
  ( \frac 5 2 )} \Big]^+$ \\
  & $2 \times \left( \frac 1 2 , \frac 3 2 \right)^+
  _{\vphantom\dagger}$
  & $\left( \frac 1 2 , \frac 3 2 , \frac 5 2 \right)^+$ \\
  \hline
  $\Sigma^+ (1D)$ & $\Big[ \tilde P_{\frac 1 2 \, D}^{( \frac 3 2 ),
  (\frac 5 2 )} \Big]^+, \ \Big[ \tilde P_{\frac 1 2 \, D}^{\prime
  \, ( \frac 3 2 ), (\frac 5 2)} \Big]^+$ &
  $\Big[ P_{\frac 3 2 \, D}^{( \frac 1 2 ), ( \frac 3 2 ),
  ( \frac 5 2 ), ( \frac 7 2 )} \Big]^+$ \\
  & $2 \times \left( \frac 3 2 , \frac 5 2 \right)^-
  _{\vphantom\dagger}$
  & $\left( \frac 1 2 , \frac 3 2 , \frac 5 2 , \frac 7 2 \right)^-$
  \\
  \hline
  $\Pi^+ (1P)$ \& \ & $\Big[ \tilde P_{\frac 1 2 \, P}^{( \frac 1 2 ),
  (\frac 3 2 )} \Big]^-, \ \Big[ \tilde P_{\frac 1 2 \, P}^{\prime
  \, ( \frac 1 2 ), (\frac 3 2)} \Big]^-$ &
  $\Big[ P_{\frac 3 2 \, P}^{( \frac 1 2 ), ( \frac 3 2 ),
  ( \frac 5 2 )} \Big]^-$ \\
  $\Sigma^- (1P)$ & $2 \times \left( \frac 1 2 ,
  \frac 3 2 \right)^-_{\vphantom\dagger}$
  & $\left( \frac 1 2 , \frac 3 2 , \frac 5 2 \right)^-$ \\
  \hline
  $\Pi^- (1P)$ & \multicolumn{2}{c}{Same as
  $\Sigma^+ (1P)_{\vphantom{\big[ }}$} \\
  \hline
  $\Sigma^- (1S)$ & $\tilde P_{\frac 1 2 \, S}^{( \frac 1 2 )-}$,
  $\tilde P_{\frac 1 2 \, S}^{\prime \, ( \frac 1 2 )-}$ &
  $P_{\frac 3 2 \, S}^{( \frac 3 2 ) -}$ \\
  & $2 \times {\frac 1 2}^+_{\vphantom\dagger}$ & ${\frac 3 2}^+$ \\
  \hline
  $\Pi^+ (1D)$ & $\Big[ \tilde P_{\frac 1 2 \, D}^{( \frac 3 2 ),
  (\frac 5 2 )} \Big]^-, \ \Big[ \tilde P_{\frac 1 2 \, D}^{\prime
  \, ( \frac 3 2 ), (\frac 5 2)} \Big]^-$ &
  $\Big[ P_{\frac 3 2 \, D}^{( \frac 1 2 ), ( \frac 3 2 ),
  ( \frac 5 2 ), ( \frac 7 2 )} \Big]^-$ \\
  & $2 \times \left( \frac 3 2 , \frac 5 2 \right)^+
  _{\vphantom\dagger}$
  & $\left( \frac 1 2 , \frac 3 2 , \frac 5 2 , \frac 7 2 \right)^+$
  \\
\hline\hline
\end{tabular}
\end{table*}

\section{Comparison to Experiment}
\label{sec:Pheno}

\subsection{Exotic Candidates of Known Quantum Numbers}

Despite the large number of exotic candidates observed, rather few
have experimentally well-determined $J^{PC}$ (or $J^{PG}$)
values~\cite{Lebed:2016hpi}.  Moreover, none of those yet seen carry
exotic $\qq$-meson or $qqq$-baryon quantum numbers,\footnote{A
possible exception is the yet-unobserved neutral partner to the
$Z_c^+(4240)$~\cite{Aaij:2014jqa}, which would have $J^{PC} \! = \!
0^{--}.$} so that the known candidates can actually be described as
``cryptoexotic.''  In large part, this self-selection of quantum
numbers arises from constraints imposed by the production modes and
decay channels most easily accessible to experiment.  For example, the
$1^{--}$ channel is especially well studied because the initial-state
radiation (ISR) process $e^+ e^- \to \gamma^{\vphantom{+}}_{\rm ISR}
Y$ produces only states $Y$ with $J^{PC} \! = \!  1^{--}$.  The exotic
candidates with measured quantum numbers (including ``favored''
values) are listed in Table~\ref{Table:JPC}.  The $J^{PC}$ quantum
numbers are also assumed known for the yet-unseen neutral isospin
partners of observed charged states such as $Z_c^+(4430)$.
\begin{table*}
  \caption{Exotic candidates with experimentally determined $J^{PC}$
  quantum numbers (both umabiguous and ``favored'').  All are $c\bar
  c$-containing states, except for those carrying a $b$ subscript,
  which are $b\bar b$-containing states.}
\label{Table:JPC}
\begin{tabular}{l|l}
  \hline\hline
$0^{++^{\vphantom\dagger}}$ & $X(3915)$, $X(4500)$, $X(4700)$ \\
$0^{--}$ & $Z_c^0(4240)$ \\
$1^{--}$ & $Y(4008)$, $Y(4220)$, $Y(4260)$, $Y(4360)$, $Y(4390)$,
$X(4630)$, $Y(4660)$, $Y_b(10888)$ \\
$1^{++}$ & $X(3872)$, $Y(4140)$, $Y(4274)$ \\
$1^{+-}$ & $Z_c^0(3900)$, $Z_c^0(4200)$, $Z_c^0(4430)$,
$Z_b^0(10610)$, $Z_b^0(10650)$ \\
${\frac 3 2}^\pm_{\vphantom\dagger}$, ${\frac 5 2}^\mp$ & $P_c(4380)$,
$P_c(4450)$ \\
  \hline\hline
\end{tabular}
\end{table*}

In comparison, Table~\ref{Table:states} exhibits 5 $0^{++}$ states, 3
$0^{--}$ states, 10 $1^{--}$ states, 5 $1^{++}$ states, and 8 $1^{+-}$
states.  Table~\ref{Table:states2} exhibits multiple spin-$\frac 3 2$
and spin-$\frac 5 2$ states of either parity.  The known exotic
candidates do not exhaust the lowest multiplets ($n \!  = \! 1$ and $L
\! \le \! 2$).  Nor does this counting take into account the likely
possibility that exotics like $Y(4140)$ decaying into $J/\psi \,
\phi$ are $c\bar c s\bar s$ states, which frees up even more possible
$c\bar c \qq$ states from Table~\ref{Table:states} for identification
with the observed exotic candidates.  To proceed further, we next
address whether heavy-quark spin symmetry, or whether selection rules
(either exact or obtained from the BO potentials), can be used to
constrain the possible identifications of states.

\subsection{Heavy-Quark Spin Symmetry}

Evidence for whether heavy-quark spin symmetry imposes strong
constraints on the exotic candidates is not without ambiguity.  If
$s_\QQ$ is a good quantum number for the exotics, then they should
decay exclusively to $\psi (\Upsilon)$ or $\chi_Q$ if $s_\QQ \! = \!
1$, and exclusively to $\eta_Q$ or $h_Q$ if $s_\QQ \! = \! 0$.

No exotic candidate has yet been observed to decay to $\eta_Q$.  In
the case of $X(3872) \! \to \! \eta_c$, the Particle Data
Group~\cite{Olive:2016xmw} presents an upper bound.  However, the
reconstruction of $\eta_Q$ states tends to be more difficult than that
for $\psi (\Upsilon)$, $\chi_Q$, or even $h_Q$ states, so it is
difficult to draw any definite conclusion in this case.

In the $c\bar c$ sector, the charmonium decays of most of the exotic
candidates proceed exclusively through $J/\psi$ or $\psi(2S)$, while a
few (such as $Z_c^+(4250)$~\cite{Mizuk:2008me}) have been seen only
with $\chi_c$ decays.  The charmonium decays of the charged
$Z_c^+(3900)$ have so far only been seen in the $J/\psi$ channel,
while those of the $Z_c^+(4020)$ have only been seen in the $h_c$
channel~\cite{Ablikim:2013wzq,Collaboration:2017njt}, suggesting
strong support for the exotic candidates appearing in eigenstates of
heavy-quark spin.

However, interesting conflicting signals occur in the region of the
$Y(4260)$, which increasingly appears to be not a single state but
several closely spaced ones~\cite{Ablikim:2016qzw,BESIII:2016adj}.  At
a bare minimum, these states appear to be the $Y(4360)$ decaying to
$\psi$, the $Y(4390)$ decaying to $h_c$, and the
$Y(4220)$,\footnote{Called $Y(4230)$ in Ref.~\cite{Lebed:2016hpi} and
elsewhere.} originally seen to decay to $\chi_{c0} \,
\omega$~\cite{Ablikim:2014qwy}, but also appearing in $h_c \pi \pi$.
The first two of these states are of course consistent with being
$s_\QQ$ eigenstates, but the latter, should it persist as a single
state, is not.

The evidence for heavy-quark spin symmetry in the $b\bar b$ sector is
much more ambiguous.  There, all the known candidate exotics
[$Y_b(10888)$, $Z_b(10610)$, $Z_b(10650)$] possess substantial decay
branching fractions into both $\Upsilon$ and
$h_b$,\footnote{See~\cite{Lebed:2016hpi} for collected experimental
references.} suggesting either that heavy-quark spin symmetry is
actually strongly violated in the decays,\footnote{Such strong
violations seem unlikely, particularly in the $b$ system, since their
amplitudes are suppressed by $\Lambda_{\rm QCD}/m_Q$.} or simply that
the resonances produced are mixtures of heavy-quark spin eigenstates.
For example, $Y_b(10888)$ might not be the state $\tilde X^{(0) ++}_{0
\, P}$ or $\tilde X^{\prime \, (0) ++}_{0 \, P}$, which are pure
$s_\QQ
\!  = \! 0$ and $s_\QQ \! = \! 1$, respectively, but rather a pure
diquark-spin eigenstate $X^{(0) ++}_{0 \, P}$ or $X^{\prime \, (0)
++}_{0 \, P}$ [Eqs.~(\ref{eq:Swavediquark})--(\ref{eq:HQbasis})].  The
latter possibility appears perhaps more plausible since the diquarks
are more compact due to the presence of the heavier $b$ quarks, but
drawing such a conclusion must await a more detailed dynamical study.

A similar situation of a given exotic state not corresponding to an
eigenstate of a single $s_\QQ$ value arises if the heavy exotics are
molecules of hadrons in their separate spin eigenstates ({\it e.g.}, a
$1^{-+}$ $\bar B^* \! B^*$ state with no admixture of $\bar B B^* \!
+ \! \bar B^* \!  B$, where $B, B^*$ has $J^P \! = \! 0^-, 1^-$,
respectively), a fact that is very well appreciated in the
construction of such models~\cite{Guo:2017jvc,Cleven:2015era}.  In
either the molecular or the diquark-antidiquark spin-eigenstate limit,
the spins of the heavy quarks are correlated not to each other (except
in the composition of the overall state $J^P$), but to the spins of
the corresponding light quarks with which they form hadron subunits,
either $(q Q) \! + \! (\bar q \bar Q)$ diquarks or $(\bar q Q) \! + \!
(\bar Q q)$ hadrons.  In both cases, the full state need not be an
eigenstate of a single $s_\QQ$ eigenvalue.

Inasmuch as heavy-quark spin symmetry does in fact hold for the
exotics, Tables~\ref{Table:states}--\ref{Table:states2} are presented
in a manner conducive to enumerating them.  Specifically, with
reference to Eqs.~(\ref{eq:SwaveQQ}), (\ref{eq:HQbasis}), and
(\ref{eq:PentaHQS}), the leftmost entries on each line
(Table~\ref{Table:states}: $\tilde X_0, \tilde Z$;
Table~\ref{Table:states2}: $\tilde P_{\frac 1 2}$) have $s_\QQ
\! = \!  0$, while the rightmost entries (Table~\ref{Table:states}:
$\tilde Z^\prime, X_1, X_2$; Table~\ref{Table:states2}: $\tilde
P^\prime_{\frac 1 2}, P_{\frac 3 2}$) have $s_\QQ \! = \! 1$.  Then,
resolving into the categories ($\{ s_\QQ \! = \! 0 \} \! + \!
\{ s_\QQ \! = \!  1 \}$), Table~\ref{Table:states} exhibits (2+3)
$0^{++}$ states, (2+1) $0^{--}$ states, (4+6) $1^{--}$ states, (1+4)
$1^{++}$ states, and (3+5) $1^{+-}$ states.  Table~\ref{Table:states2}
exhibits (3+7) ${\frac 3 2}^+$, (2+5) ${\frac 3 2}^-$, (1+4) ${\frac 5
2}^+$, and (1+3) ${\frac 5 2}^-$ states.

Interestingly, the pentaquark candidates $P_c(4380)$ and $P_c(4450)$
both decay to $J/\psi \, p$, and their close spacing in mass combined
with the large width for $P_c(4380)$ and small width for $P_c(4450)$
suggests---at this stage---that $P_c(4380)$ is the highest
$\Sigma^+(1S)$ state $P^{(\frac 3 2)+}_{\frac 3 2 \, S}$ ($J^P =
{\frac 3 2}^-$), while $P_c(4450)$ is the unique ${\frac 5 2}^+$ state
in the lowest multiplets,\footnote{Similar reasoning in a Hamiltonian
formalism led to the same $J^P$ identification of the $P_c$ states in
the triquark-diquark model of Ref.~\cite{Zhu:2015bba}.} namely, the
$\Sigma^+(1P)$ state $P^{(\frac 5 2)+}_{\frac 3 2 \, P}$.

\subsection{Selection Rules}

Selection rules for strong decays of exotics in the BO approach were
first developed in Ref.~\cite{Braaten:2014ita} and applied
systematically in Ref.~\cite{Braaten:2014qka}.  These selection rules
fall into three types.  The first is overall conservation of $J^{PC}$
for the process, which is exact in strong interactions.  Assuming that
the initial and final $\QQ$-containing states have quantum numbers
$J_i^{P_i C_i}$ and $J_{\! f}^{P_{\! f} C_{\! f}}$, respectively, and
a single hadron with quantum numbers $j^{pc}$ is emitted with orbital
angular momentum $\ell$ relative to the final heavy state, one
immediately has the selection rules
\begin{eqnarray}
P_i & = & P_{\! f} \, p \, (-1)^\ell \, , \nonumber \\
C_i & = & C_{\! f} \, c \, \nonumber \\
\bm{J}_i & = & \bm{J}_{\! f} + \bm{j} + \bm{\ell} \, .
\label{eq:JPCCond}
\end{eqnarray} 
The $C$ eigenvalues here refer of course only to the neutral members
of each isospin multiplet; if transitions involving charged states are
considered, then $G$-parity conservation, with $G \! = \! C (-1)^I$
for each state, must be imposed.  Selection rules in this class are
never violated, assuming only that the decays are pure QCD processes.

The second type of selection rule in Ref.~\cite{Braaten:2014qka}
references the approximate conservation of heavy-quark spin symmetry.
We have already discussed in the previous subsection how well this
symmetry is upheld in observed processes.

The third type of selection rule in Ref.~\cite{Braaten:2014qka} uses
the BO approximation in a fundamental way: Under the assumption that
the light d.o.f.\ adjust much more quickly in a physical process than
the heavy $\QQ$ pair, the decay to a conventional $\QQ$ state occurs
through a rapid transition from the initial BO configuration to the
final one plus a light hadron, leaving the separation and orientation
of the $\QQ$ pair nearly unchanged.  In fact, the heavy-quark
spin-symmetry limit is implicit in this approximation.  The
conservation of angular momentum then reads
\begin{equation}
\bm{J}_{{\rm light},i} = \bm{J}_{{\rm light},f} + \bm{j} + \bm{\ell}
\, ,
\end{equation}
and since $\bm{J}_{{\rm light},i} \! = \! \bm{L}_i + \bm{s}_\qq$ (with
the replacement $\bm{s}_\qq \! \to \! \bm{s}_\B$ for pentaquarks)
while $\bm{J}_{{\rm light},f}$ (being the light d.o.f.\ angular
momentum of a conventional $\QQ$ state) contains no valence light
quarks and hence equals $\bm{L}_{\! f}$,\footnote{A distinct
$\bm{s}_\qq$ factor appears as a component of $\bm{J}_{{\rm light},f}$
if it is also a multiquark state.} we have
\begin{equation}
\bm{L}_i = \bm{L}_{\! f} + \bm{j} - \bm{s}_\qq + \bm{\ell} \, .
\end{equation}
Dotting with $\hat{\bm{r}}$ gives
\begin{equation} \label{eq:LambdaVec}
\lambda_i = \lambda_{\! f} + \hat{\bm{r}} \cdot \left( \bm{j} -
\bm{s}_\qq + \bm{\ell} \right) \, .
\end{equation}
This expression differs from Eq.~(28) in Ref.~\cite{Braaten:2014qka}
by the extra factor $-\bm{s}_\qq$ on the right-hand side, and arises
as the result of our choice not to include $\bm{s}_\qq$ in
$\bm{J}_{{\rm light},i}$.  The triangle rule for the transition in the
BO approximation reads
\begin{equation}
\left| \lambda_i - \lambda_{\! f} \right| \le j + s_\qq + \ell
\, .
\end{equation}
Again, since the final state is taken to be a conventional $\QQ$ state
with BO potential $\Sigma^+_g$, then $\lambda_{\! f} \! = \! 0$, and
thus:
\begin{equation} \label{eq:LambdaCond}
\Lambda_i \le j + s_\qq + \ell \, ,
\end{equation}
with $s_\qq \to s_\B$ in the pentaquark case.  It is interesting to
consider the limit discussed in Sec.~\ref{sec:BO} in which the
light-quark d.o.f.\ spin $\bm{s}_\qq$ also remains fixed ({\it i.e.},
for $b\bar b c\bar c$ tetraquarks, since $m_b \! \gg \! m_c
\!  \gg \! \Lambda_{\rm QCD}$).  Then, assuming the light final-state
hadron contains no internal orbital excitation, one has $\bm{j} \! =
\! \bm{s}_\qq$, and hence from Eq.~(\ref{eq:LambdaVec}) follows the
simple result $\Lambda_i \! \le \! \ell$: States in $\Sigma, \Pi,
\ldots$ BO potentials in this limit only decay to light hadrons in at
least $S,P, \ldots$ relative partial waves, respectively.  In the
light-quark case, however, only the looser constraint
Eq.~(\ref{eq:LambdaCond}) applies.

The discrete BO eigenvalues also provide approximate selection rules.
Following the analysis in Sec.~\ref{sec:BO}, the reflection parity
$R_{\rm light}$ through any plane containing the $\QQ$ axis
$\hat{\bm{r}}$ acts upon the light hadron as a product of $P_{\rm
light}$ (which introduces a factor of its intrinsic parity $p$ as well
as a factor $(-1)^\ell$ from its relative motion with respect to the
final heavy state) and a rotation by $\pi$ radians about the normal to
the reflection plane (which introduces an extra phase exp($i\pi
\hat{\bm{r}} \! \cdot \! \bm{s}_\qq$) in the initial state and
exp[$i\pi \hat{\bm{r}} \! \cdot \! ( \bm{j} \! + \!
\bm{\ell} )$] in the final state).  According to
Eq.~(\ref{eq:LambdaVec}), the difference of these phases is just
exp[$i\pi (\lambda_i - \lambda_{\! f})$], which can be written as
$(-1)^{\Lambda_i - \Lambda_{\! f}}$ since both $\lambda$'s are
integers.  In total, we have
\begin{equation} \label{eq:EpsCond}
\epsilon_i = \epsilon_{\! f} \, p \, (-1)^\ell
(-1)^{\Lambda_i - \Lambda_{\! f}} \, , \ \, {\rm or} \
\rho_i = \rho_{\! f} \, p \, (-1)^\ell \, .
\end{equation}
We note that no restriction to $\Lambda_i \! = \Lambda_{\! f} \! = \!
0$ is required, in contrast to Eq.~(31) of
Ref.~\cite{Braaten:2014qka}.

Lastly in the tetraquark case, for which charge conjugation symmetry
is relevant, the BO approximation $(CP)_{\rm light}$ quantum number
$\eta$ provides a selection rule (Eq.~(30) of~\cite{Braaten:2014qka}):
\begin{equation} \label{eq:EtaCond}
\eta_i = \eta_f \, c p \, (-1)^\ell \, , \ \, {\rm or} \
\kappa_i = \kappa_{\! f} c \, .
\end{equation}

The most incisive phenomenological tests of the exact selection rules
Eqs.~(\ref{eq:JPCCond}) and the BO approximation selection rules
Eqs.~(\ref{eq:LambdaCond}), (\ref{eq:EpsCond}), (\ref{eq:EtaCond}) are
decays to conventional $\QQ$ states ($\epsilon_{\!  f} \! = \! \eta_f
\! = \! +$, $\lambda_{\! f} \!  = \! 0$) that produce a single light
hadron.  The decays of this type thus far observed are the emission of
a single light vector particle ($j^{pc} \! = \!  1^{--}$, $\ell \! =
\! 0$) such as $\rho$, $\omega$, or $\phi$, and the emission of a
single charged pion ($j^{pg} \! = \!  0^{--}$, $\ell \! = \! 0$).  In
the latter case, since the conventional $\QQ$ states are isosinglets,
one has an isotriplet exotic decaying to a single pion, in which case
the $(-1)^I \! = \!  -1$ factors in the definition of $G$ parity
cancel between the initial and final state, thus reducing $G$-parity
conservation condition to the $C$-parity conservation condition
[Eq.~(\ref{eq:JPCCond})] for the corresponding $\pi^0$ process, $c \!
= \! +$ and $C_i \! = \! C_{\! f}$.

Let us first consider the pionic decay.  The role of $\pi$ as a
Nambu-Goldstone boson of chiral symmetry breaking suggests it to be
emitted predominantly in a $P$-wave ($\ell \! = \! 1$).  However,
$Z_c^+(3900)$ has been experimentally determined to be a $1^+$
state~\cite{Collaboration:2017njt}, and therefore the observed decay
$Z_c^+(3900) \! \to \!  J/\psi \, \pi^+$ to the $1^{--}$ $J/\psi$
requires $\ell$ to be even for this decay.  Presumably, the $S$ wave
must dominate this particular process; if this result remains true for
the other single-pion emission processes, then the selection rules
reduce to
\begin{eqnarray}
  & & P_i = -P_{\! f}, \ C_i = C_{\! f}, \ J_i = J_{\! f}, \nonumber
  \\
  & & \Lambda_i \le s_\qq, \ \epsilon_i =
  (-1)^{\Lambda_i + 1}, \ \eta_i = - \, , \nonumber \\
  & & [ \,\rho_i \kappa_i = -+ \, ] \, .
\end{eqnarray}
In particular, only $u$ BO potentials for the initial states are
represented.  Since $J/\psi$ is $1^{--}$ and $\pi^0$ is $0^{-+}$,
$Z_c^0(3900)$ is therefore $1^{+-}$, and a glance at
Table~\ref{Table:states} shows 3 $s_\QQ \! = \! 1$ candidates in a
$\rho_i \kappa_i = -+$ BO potential with these quantum numbers,
namely, $\tilde X^{\prime \, (1)-+}_{0 \, P}$, $X^{(1)-+}_{1 \, P}$,
and $X^{(1)-+}_{2 \, P}$ in the mixed $\Pi^+_u(1P)$-$\Sigma^-_u(1P)$
BO potential.  Should the $Z_c^+(4020)$ (which decays to $h_c$) also
be confirmed as a $1^+$ state, its natural identification would be as
the $s_\QQ \! = \! 0$ state $\tilde X^{(1)-+}_{0 \, P}$ in the same BO
potential.  The $Z_c^+(4200)$ and $Z_c^+(4430)$ can be analyzed
similarly, but whether they are the other two
$\Pi^+_u(1P)$-$\Sigma^-_u(1P)$ states, or belong to either a higher $n
\! = \! 1$ BO potential or the $n \! = \! 2$ band, requires a more
detailed study.  The $Z_c^0(4240)$, should its $0^{--}$ quantum
numbers be unambiguously confirmed, is more problematic because it
does not fit into the $\Pi^+_u(1P)$-$\Sigma^-_u(1P)$ BO potential with
an $S$-wave pion coupling, but with a $P$-wave decay it could be the
state $X^{(0)++}_{1 \, P}$ in $\Sigma_g^+(1P)$.

Turning now to the single light-vector decays and assuming $S$-wave
decays, the selection rules reduce to
\begin{eqnarray}
  & & P_i = -P_{\! f} , \ C_i = -C_{\! f}, \ J_i \in
  \{ J_{\! f}, \, J_{\! f} \pm 1 \}, \nonumber \\
  & & \Lambda_i \le 1 + s_\qq, \ \epsilon_i =
  (-1)^{\Lambda_i + 1}, \ \eta_i = + \, , \nonumber \\
  & & [ \,\rho_i \kappa_i = -- \, ] \, .
\end{eqnarray}
In particular, only $g$ BO potentials for the initial states are
represented.  A quick glance at Table~\ref{Table:states} shows that no
$\rho_i \kappa_i = --$ potentials are expected among the lowest
multiplets, which creates a real problem for this classification.  It
could be resolved in several ways: First, the BO approximation for
exotic states might simply not work because the physical values $m_Q
\! = \!  m_c, m_b$ are not large enough; however, inasmuch as the
approximation becomes exact for $m_Q \! \to \! \infty$, it would be
peculiar for the classification to fail for {\em every\/}\
light-vector decay mode but still work for the single-pion decay
modes.  Second, the BO approximation is expected to fail in for states
in the vicinity of two-hadron thresholds, at which point avoided
energy-level crossings must be taken into account by means of a
coupled-channel analysis, as discussed in~\cite{Braaten:2014qka} or
implemented via the Feshbach mechanism in~\cite{Esposito:2016itg};
while this observation is certainly true and will have to be
implemented in a fully complete model, not every exotic candidate
(even restricting to ones decaying to light vectors) is especially
close to such a threshold.  In either of these first two scenarios,
the conservation of the BO quantum numbers can be violated in decay
transitions.  Third, the light vectors might (for unknown reasons)
couple predominantly to a $P$ wave, in which case one finds $\rho_i
\kappa_i = +-$; while the $\Pi^-_u(1P)$ potential fits this category
and indeed produces $1^{--}$ states, it produces neither $0^{++}$ nor
$1^{++}$ states.

A fourth option is that the listing of the lowest BO potentials for
$\delta$-$\bar \delta$ given in Table~\ref{Table:states} is
incomplete.  One particularly economical solution is to suppose that
the potentials $\Pi^+_g(1P)$ and $\Pi^-_g(1P)$ are among the lowest.
Following the comments below Eq.~(\ref{eq:FinalPC}), the listing of
states for $\Pi^+_g(1P)$ looks exactly like that for
$\Pi^+_u(1P)$-$\Sigma^-_u(1P)$, except that all final superscripts,
$\kappa$ and $C$, flip sign.  Then several $0^{++}$ and $1^{++}$
states naturally appear [and, according to our previous discussion, it
matches the quantum numbers of states in---and potentially mixes
with---$\Sigma^-_g(1P)$].  Likewise, the listing of states for
$\Pi^-_g(1P)$ [which may mix with $\Sigma^+_g(1P)$] looks exactly like
that for $\Pi^-_u(1P)$ except for the flip of $\kappa$ and $C$, which
naturally produces multiple $1^{--}$ states (as well as another option
for a $0^{--}$ state).  For completeness, these additional multiplets
are listed in Table~\ref{Table:states3}.  Whether this resolution is
reasonable of course depends upon the true ordering of $\delta$-$\bar
\delta$ BO potentials, which presumably can be decided by lattice
simulations.  For example, simulations such as those described for
$b\bar b u\bar d$ in Ref.~\cite{Peters:2017hon} will be quite
valuable.
\begin{table*}
  \caption{Quantum numbers for possible additional low-lying
  tetraquark states in the dynamical diquark picture, as suggested by
  the Born-Oppenheimer selection rules for light-vector decays.  The
  notation is the same as in Table~\ref{Table:states}.}
\label{Table:states3}
\begin{tabular}{c|c|c|c}
  \hline\hline
  BO potential & \multicolumn{3}{c}{State notation} \\
  & \multicolumn{3}{c}{State $J^{PC}$} \\
  \hline
  $\Pi^+_g(1P)$ & $\tilde X_{0 \, P}^{(1) --}$ &
  $[\tilde Z_P^{\bm{(0)}, (1), \bm{(2)}}]^{--}$,
  $[\tilde Z_P^{\prime \, \bm{(0)}, (1), \bm{(2)}}]^{--}$ &
  $\tilde X_{0 \, P}^{\prime \, (1) --}$, \
  $[X_{1 \, P}^{(0), (1), (2)}]^{--}$, \
  $[X_{2 \, P}^{(1), (2), (3)}]^{--}$ \\
  & $1^{++}$ & $2 \times (\bm{0},1,\bm{2})^{+-}$ &
  $[ 1, \ (0,1,2), \ (1,2,3) ]^{++}$ \\
  \hline
  $\Pi^-_g(1P)$ & $\tilde X_{0 \, P}^{(1) ++}$ &
  $[\tilde Z_P^{(0), \bm{(1)}, (2)}]^{++}$,
  $[\tilde Z_P^{\prime \, (0), \bm{(1)}, (2)}]^{++}$ &
  $\tilde X_{0 \, P}^{\prime \, (1) ++}$, \
  $[X_{1 \, P}^{\bm{(0)}, (1), (2)}]^{++}$, \
  $[X_{2 \, P}^{(1), (2), (3)}]^{++}$ \\
  & $1^{--}$ & $2 \times (0,\bm{1},2)^{-+}$ &
  $[ 1, \ (\bm{0},1,2), \ (1,2,3) ]^{--}$ \\
\hline\hline
\end{tabular}
\end{table*}

Finally, the $\bar \theta$-$\delta$ BO states have the same selection
rules, excluding those for $C$ and $\eta$~[Eq.~(\ref{eq:EtaCond})],
while $s_\qq$ is replaced by $s_\B$, which we take to have its minimal
value, $\frac 1 2$.  Assuming for now that we are only interested to
decays into $J/\psi$ ($J_{\!  f}^{P_{\! f}} \! = \! 1^-$) and nucleons
($j^p \! = \! {\frac 1 2}^+$), the selection rules read
\begin{eqnarray} \label{eq:PentaConst}
& & P_i = \rho_i (-1)^{L+1} = P_{\! f} \, p \, (-1)^\ell =
(-1)^{\ell + 1}, \ J_i \le \h3 + \ell \, , \nonumber \\
& & \Lambda_i \le 1 + \ell, \ \epsilon_i = (-1)^{\Lambda_i + \ell} \,
, \ [ \, \rho_i = (-1)^\ell \, ] \, .
\end{eqnarray}
Since the two observed $P_c$ states are found to have opposite
parities, the first string of equations in (\ref{eq:PentaConst}) shows
that opposite parities of $\ell$ are required to accommodate them.  In
addition, the $J_i$ triangle rule requires $\ell \! \ge \! 1$ for the
spin-$\frac 5 2$ state.  Moreover, substituting the final equality in
(\ref{eq:PentaConst}) into its first equation shows that $L$ must be
even for the potentials producing each state.  The $J_i^{P_i}$ option
${\frac 3 2}^- \! , \, {\frac 5 2}^+$ for the $P_c$ states [which
suits the broad width of the $P_c(4380)$ and narrow width of the
$P_c(4450)$, as discussed in the previous subsection] corresponds to
$\ell \! = \! 0, 1$, respectively, and is accommodated most naturally
by the pair $\Sigma^+(1S)$: $P^{({\frac 3 2}) +}_{{\frac 3 2} \, S}$,
and $\Pi^+(1D)$ [or $\Sigma^-(1D)$, if also pre\-sent]: $\tilde
P^{\prime \, ({\frac 5 2}) - \!}_{{\frac 1 2} \, D}$ or $P^{({\frac 5
2}) -}_{{\frac 3 2} \, D}$.  Alternately, if the $P_c$ states are
found to be ${\frac 3 2}^+ \! , \, {\frac 5 2}^-$, then the pair
$\Sigma^-(1S)$: $P^{(\frac 3 2) -}_{{\frac 3 2} \, S}$, and
$\Sigma^+(1D)$ [or $\Pi^-(1D)$, if also present]: $\tilde P^{(\frac 5
2) +}_{{\frac 1 2} \, D}$ or $P^{({\frac 5 2}) +}_{{\frac 3 2} \, D}$
works.

In summary, the most natural BO potentials for accommodating known
tetraquark candidates appear to be $\Pi^+_u(1P)$-$\Sigma^-_u(1P)$ for
those appearing in single-pion decays, $\Pi^+_g(1P)$-$\Sigma^-_g(1P)$
and $\Pi^-_g(1P)$-$\Sigma^+_g(1P)$ for those appearing in
single-vector decays.  The most natural BO potentials for
accommodating the known pentaquark candidates, depending upon the
final parity assignments, are $\Sigma^+(1S)$ and $\Pi^+(1D)$, or
$\Sigma^-(1S)$ and $\Sigma^+(1D)$.

\section{Conclusions}
\label{sec:Concl}

In this paper we developed the spectroscopy of states in the dynamical
diquark picture, both for diquark-antidiquark tetraquark states and
for triquark-diquark pentaquark states, within the context of the
Born-Oppenheimer (BO) approximation.  The first step was the
group-theoretical exercise of relating the diquark-spin basis to the
basis of states of well-defined heavy-quark spin, in which $P$ and $C$
quantum numbers are most easily determined.

Next, the BO approximation and potentials were briefly reviewed, the
quantum numbers of states within these potentials were determined, and
a compact notation for the states was introduced.  The lowest BO
potentials were identified by supposing that the lowest potentials
obtained in lattice QCD simulations for hybrid mesons in the BO
approximation hold also for diquark-antidiquark and triquark-diquark
systems.  Then the lowest expected multiplets for states in the
diquark-antidiquark system were collected in Table~\ref{Table:states},
and in the triquark-diquark system in Table~\ref{Table:states2}.
 
We then turned to the question of comparison with the set of exotic
candidates with experimentally observed quantum numbers, and found
that all of these states could be accommodated, taking into account
just their $J^{PC}$ quantum numbers.  We then developed selection
rules---some exact and some relying upon the BO approximation---for
decays of the exotic states, and carefully examined the constraints
thus obtained using known decay channels (via a single pion or a
single light vector meson).  We found that the observed tetraquark
states decaying through pions and the pentaquark states could still be
accommodated, but if the BO selection rules must hold in their
strictest form, then the single-vector decays appear to demand the
introduction of additional low-lying BO potentials
[Table~\ref{Table:states3}] beyond the ones appearing in hybrid
lattice calculations.

The next steps of this study point in many different directions.
First, it is important to keep track of the latest discoveries in the
exotic sector, to see whether newly discovered states or old states
with newly determined quantum numbers continue to fit into the BO
paradigm.  Second, lattice simulations of the lowest BO potentials
that include nontrivial light-quark spin or isospin will be essential
in firming up the identification of the known exotics with particular
states and determining whether the BO selection rules actually hold in
all instances.  Third, particular functional forms for the potentials
inspired by lattice results or models can be introduced, and the
corresponding Schr\"{o}dinger equations solved, in order to obtain
predictions for the specific mass spectrum of the states.  Fourth, the
very interesting question of how coupled-channel effects with hadronic
thresholds modify these predictions must be addressed, as it cannot
simply be an accident that many of the exotic candidates lie so close
in mass to such thresholds (especially $m_{X(3872)} \! - \! m_{D^{*0}}
\! - \! m_{D^0} \! = \! +0.01 \! \pm \!
0.18$~MeV~\cite{Olive:2016xmw}).

An ambitious program of calculations within not only the diquark model
but molecular models as well, combined with the steady rate of new
experimental and lattice simulation developments, will lead to a much
richer and clearer understanding of these novel hadrons.

\begin{acknowledgments}
  \vspace{-2ex} This work was supported by the National Science
  Foundation under Grant No.\ PHY-1403891.
\end{acknowledgments}


\end{document}